\documentclass[sigconf,screen,nonacm]{acmart}

\AtBeginDocument{%
  }

\usepackage{enumitem}

\begin{document}

\title{X-VC: Zero-shot Streaming Voice Conversion in Codec Space}

\author{Qixi Zheng}
\affiliation{
  \institution{Shanghai Jiao Tong University}
  \city{Shanghai}
  \country{China}
}
\email{jerrister.150329@sjtu.edu.cn}

\author{Yuxiang Zhao}
\affiliation{
  \institution{Shanghai Jiao Tong University}
  \city{Shanghai}
  \country{China}
}

\author{Tianrui Wang}
\affiliation{
  \institution{Tianjin University}
  \city{Tianjin}
  \country{China}
}

\author{Wenxi Chen}
\affiliation{%
  \institution{Shanghai Jiao Tong University, Shanghai Innovation Institute}
  \city{Shanghai}
  \country{China}
}

\author{Kele Xu}
\affiliation{%
  \institution{State Key Laboratory of Complex \& Critical Software Environment}
  \city{Changsha}
  \country{China}
}

\author{Yikang Li}
\affiliation{
  \institution{Shanghai Innovation Institute}
  \city{Shanghai}
  \country{China}
}

\author{Qinyuan Cheng}
\affiliation{
  \institution{Fudan University, \\ Shanghai Innovation Institute}
  \city{Shanghai}
  \country{China}
}

\author{Xipeng Qiu}
\affiliation{
  \institution{Fudan University, \\ Shanghai Innovation Institute}
  \city{Shanghai}
  \country{China}
}

\author{Kai Yu}
\affiliation{
  \institution{Shanghai Jiao Tong University}
  \city{Shanghai}
  \country{China}
}

\author{Xie Chen}
\authornote{Corresponding author.}
\affiliation{
  \institution{Shanghai Jiao Tong University, Shanghai Innovation Institute}
  \city{Shanghai}
  \country{China}
}
\email{chenxie95@sjtu.edu.cn}

\renewcommand{\shortauthors}{Zheng et al.}

\begin{abstract}
Zero-shot voice conversion (VC) aims to convert a source utterance into the voice of an unseen target speaker while preserving its linguistic content.
Although recent systems have improved conversion quality, building zero-shot VC systems for interactive scenarios remains challenging because high-fidelity speaker transfer and low-latency streaming inference are difficult to achieve simultaneously.
In this work, we present X-VC, a zero-shot streaming VC system that performs one-step conversion in the latent space of a pretrained neural codec.
X-VC uses a dual-conditioning acoustic converter that jointly models source codec latents and frame-level acoustic conditions derived from target reference speech, while injecting utterance-level target speaker information through adaptive normalization.
To reduce the mismatch between training and inference, we train the model with generated paired data and a role-assignment strategy that combines standard, reconstruction, and reversed modes.
For streaming inference, we further adopt a chunkwise inference scheme with overlap smoothing that is aligned with the segment-based training paradigm of the codec.
Experiments on Seed-TTS-Eval show that X-VC achieves the best streaming WER in both English and Chinese, strong speaker similarity in same-language and cross-lingual settings, and substantially lower offline real-time factor than the compared baselines.
These results suggest that codec-space one-step conversion is a practical approach for building high-quality low-latency zero-shot VC systems.
Our audio samples, code and checkpoints are released at \url{https://github.com/Jerrister/X-VC}.
\end{abstract}



\maketitle

\section{Introduction}

Voice conversion (VC) aims to transform a source utterance so that it sounds as if spoken by a target speaker while preserving the original linguistic content.
It has broad applications in multimedia and speech technologies, including dubbing and localization, character voice editing in films, games, and animation, personalized speech generation, and assistive communication \cite{Mohammadi2017Overview,Sisman2021Overview}.
Zero-shot or any-to-any VC further requires the model to generalize to unseen speakers without speaker-specific fine-tuning.
A practical zero-shot VC system therefore needs to preserve source linguistic content, accurately transfer target speaker characteristics, and support low-latency streaming inference in a unified framework.

Prior work often approaches zero-shot VC via an analysis and synthesis pipeline, extracting speaker-independent content from the source and combining it with target information to reconstruct the waveform \cite{Qian2019AutoVC,Wang2021VQMIVC,Lin2021FragmentVC,11209157,Li2022FreeVC,Li2021StarGANv2VC}.
While these studies demonstrate the feasibility of non-parallel zero-shot VC, recent systems highlight the need for richer target-side conditioning. For example, SEF-VC learns timbre directly from reference speech via cross-attention rather than explicit embeddings \cite{Li2024SEFVC}, and models like Vevo and vec2wav~2.0 show that timbre-dependent target conditioning substantially improves conversion quality \cite{Zhang2025Vevo,Guo2024Vec2Wav2}. Consequently, fine-grained target timbre transfer remains a central challenge in zero-shot VC, especially for unseen speakers.

To better align training supervision with the inference-time conversion scenario, recent work has also begun to leverage generated paired data.
Instead of relying only on self- or cross-reconstruction from real utterances, this paradigm constructs pseudo-parallel supervision in which the linguistic content comes from one utterance while the target timbre is provided by another.
A representative example is Seed-VC \cite{Liu2024SeedVC}, which explicitly identifies timbre leakage, insufficient timbre representation, and training--inference inconsistency as key obstacles in zero-shot VC.
By introducing content-aligned generated pairs and conditioning on the full reference speech context, Seed-VC shows that generated-data training can provide stronger supervision for learning content--timbre recombination in zero-shot conversion.
This suggests that the key issue is not whether generated data should replace conventional zero-shot VC pipelines, but how target-side information can be more effectively exploited for voice conversion.

At the same time, many real-world scenarios require low-latency or real-time conversion, which places stringent constraints on model design.
Recent streaming VC systems have shown that chunkwise or blockwise processing is a viable route to low-latency inference \cite{Yang2024StreamVC,Wang2024StreamVoice,Ning2023DualVC,Ning2024DualVC2,Zhang2025Conan,Ma2025MeanVC}.
However, maintaining strong speaker similarity, content fidelity, and naturalness under streaming constraints remains challenging, especially in zero-shot settings where the target speaker is unseen and only reference speech is available.
As a result, jointly achieving high-fidelity target timbre transfer, robust content preservation, and low-latency streaming remains a central open problem in practical zero-shot VC.

Another relevant observation comes from recent speech generation research, where codec tokens and codec latents have become increasingly attractive modeling spaces.
Modern neural codecs provide high-fidelity reconstruction and compact intermediate representations \cite{Zeghidour2021SoundStream,Defossez2022Encodec}, while codec-based generation has driven rapid progress in zero-shot speech generation \cite{Wang2023VALLE,Kharitonov2023SPEARTTS,Borsos2023SoundStorm,Ju2024NaturalSpeech3}.
This suggests that codec space is a natural option for VC as well: the conversion model can focus on transforming a reconstruction-oriented latent representation, while waveform synthesis is delegated to a pretrained codec decoder.
Therefore, the latent space of a pretrained neural codec provides a powerful and efficient interface for high-quality, low-latency voice conversion.

Yet operating in codec space does not remove the central conditioning challenge.
High-quality VC requires both frame-level acoustic cues and utterance-level speaker information.
Fine-grained target timbre is better captured by sequence-level or full-reference conditioning \cite{Lin2021FragmentVC,Li2024SEFVC,Liu2024SeedVC,Guo2024Vec2Wav2}, while utterance-level speaker representations remain important for global identity consistency \cite{Li2022FreeVC,Casanova2022YourTTS,Qin2023OpenVoice}.
In our setting, these signals are heterogeneous: source speech is represented in codec latent space, whereas the frame-level condition is represented in mel space.
A central difficulty is therefore how to model fine-grained frame-level target cues and utterance-level speaker information under streaming constraints.
More broadly, conditioning mechanisms have evolved from feature modulation schemes such as AdaIN and FiLM \cite{Huang2017AdaIN,Perez2018FiLM} to transformer-based conditional generation architectures.
Recent speech generation systems such as E2 TTS and F5-TTS show that DiT-style backbones can be effectively adapted to speech generation \cite{Eskimez2024E2TTS,Chen2025F5TTS}.
For heterogeneous inputs, architectures such as MMDiT provide a more relevant reference, where different inputs are first represented separately and then allowed to interact through joint attention \cite{Peebles2023DiT,StabilityAI2024SD3}.
These developments suggest that the key challenge is to preserve the structural differences between conditions while enabling effective interaction among them.

To address these challenges, while drawing on the above insights, we present \textbf{X-VC}, a zero-shot streaming voice conversion system in codec space.
X-VC performs one-step conversion in the latent space of a pretrained neural codec.
A dual-conditioning acoustic converter is introduced to jointly model source codec latents and reference mel-based frame-level conditions, while utterance-level speaker embeddings are injected through adaptive normalization.
High-quality generated pairs are used as supervision during training, and chunkwise inference is applied in streaming scenarios to enable low-latency conversion.
The main contributions of this work are summarized as follows:
\begin{itemize}
    \item We present a codec-space framework for zero-shot streaming voice conversion, where conversion is performed in the codec latent space rather than directly in waveform or spectrogram space.
    \item We design a dual-conditioning acoustic converter that jointly models frame-level acoustic conditions and utterance-level speaker information for codec-space latent transformation.
    \item We develop a practical training and inference framework, including generated-pair training with flexible role assignment and chunkwise streaming inference aligned with codec segmentation, improving robustness and enabling low-latency conversion, while naturally supporting cross-lingual voice conversion.
    \item We validate the proposed approach on English, Chinese, and cross-lingual settings, and show through extensive experiments and ablations the importance of dual-condition modeling and the proposed training strategy.
\end{itemize}

\section{Related Work}

\begin{figure*}[htbp]
    \centering
    \includegraphics[width=0.78\linewidth]{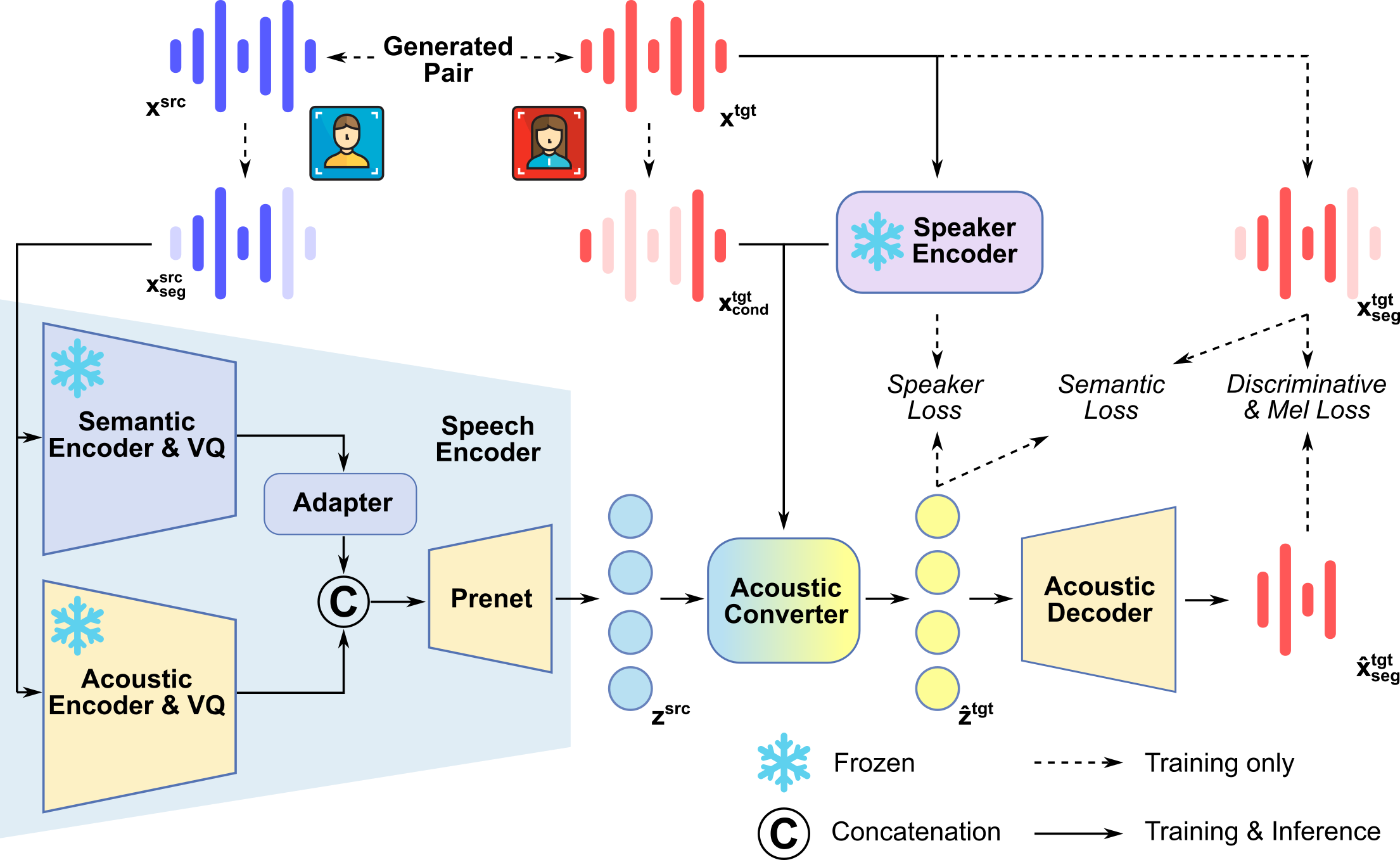}
    \caption{Overall framework of X-VC. A pretrained speech encoder maps the input waveform into latent representations, which are then transformed by the acoustic converter conditioned on target-side reference speech and finally decoded into waveform.}
    \label{fig:model_overview}
    \Description{The figure presents a pipeline consisting of multiple modules. On the left, a speech encoder includes a semantic encoder with vector quantization and an acoustic encoder with vector quantization, both marked as frozen. Their outputs are combined and passed through an adapter and a prenet module to form codec latent representations. These representations are then processed by an acoustic converter module. A speaker encoder extracts a global speaker embedding from reference audio and provides conditioning information. The converted latent representations are finally fed into an acoustic decoder to synthesize waveform audio. Colored dots represent latent sequences at different stages. Snowflake icons indicate frozen components.}
\end{figure*}

\subsection{High-Fidelity Zero-Shot VC}

Recent zero-shot VC systems increasingly explore richer target-side information to improve speaker similarity and naturalness.
SEF-VC~\cite{Li2024SEFVC} learns speaker timbre directly from reference speech through cross-attention, avoiding explicit speaker embeddings.
Vevo~\cite{Zhang2025Vevo} studies controllable zero-shot voice imitation with self-supervised disentanglement, while vec2wav~2.0~\cite{Guo2024Vec2Wav2} combines discrete content tokens with timbre-dependent features to improve conversion quality.
Compared with earlier zero-shot VC systems, these methods place more emphasis on how target-side information is represented and injected into the model.
However, retaining such rich target-side conditioning in low-latency streaming zero-shot VC remains underexplored.

\subsection{VC with Pseudo-Parallel Supervision}

Seed-VC~\cite{Liu2024SeedVC} represents a different direction, where generated paired data are introduced to better match the inference-time conversion scenario.
It constructs pseudo-parallel pairs in which the linguistic content is inherited from one utterance while the target timbre is provided by another, and trains a diffusion-based conversion model with full-reference conditioning.
This line of work demonstrates the value of generated supervision for zero-shot VC, while how to further exploit generated pairs during training has not been explicitly explored.

\subsection{Dual-Condition Modeling}

Voice conversion often requires the joint use of frame-level acoustic information and utterance-level speaker characteristics~\cite{Qian2019AutoVC}.
These two forms of conditioning are complementary, providing both time-varying and global target cues, but their differences in structure and granularity make them non-trivial to integrate within a unified model~\cite{Li2022FreeVC}.
To handle such heterogeneous conditions, models like the MMDiT~\cite{StabilityAI2024SD3} use separate projections for different modalities and enable interaction through joint attention, while adaptive layer normalization (AdaLN)~\cite{Peebles2023DiT} provides an effective way to inject global conditioning signals through feature modulation.
These developments suggest a promising direction for VC: jointly modeling frame-level and utterance-level conditions while preserving their structural differences.

\subsection{Streaming Voice Conversion}

Streaming VC focuses on reducing inference latency while preserving linguistic content and target speaker identity.
StreamVC~\cite{Yang2024StreamVC} adopts chunkwise processing for real-time low-latency conversion, Conan~\cite{Zhang2025Conan} develops a chunkwise online zero-shot adaptive VC framework, and MeanVC~\cite{Ma2025MeanVC} introduces mean flows for single-step streaming zero-shot VC.
Nevertheless, maintaining strong target speaker similarity and content fidelity in zero-shot settings remains challenging for streaming models.


\section{Methodology}
\label{sec:method}

\subsection{Overview}
\label{subsec:overview}

We formulate zero-shot voice conversion in the latent space of a pretrained neural audio codec.
Given a source utterance $\mathbf{x}^{\mathrm{src}}$ and a target-side reference utterance $\mathbf{x}^{\mathrm{tgt}}$, 
we first extract a short segment $\mathbf{x}^{\mathrm{src}}_{\mathrm{seg}}$ and encode it into codec latent representations using a pretrained speech encoder:
\begin{equation}
    \mathbf{z} = \mathcal{E}(\mathbf{x}^{\mathrm{src}}_{\mathrm{seg}}),
\end{equation}
where $\mathbf{z}$ denotes the source codec latent sequence.

For conditioning, we construct a target-side reference signal by removing the corresponding segment from the target utterance, denoted as $\mathbf{x}^{\mathrm{tgt}}_{\mathrm{cond}}$.
From this signal, we extract two complementary types of information:
a frame-level acoustic condition (e.g., mel-spectrogram) and an utterance-level speaker representation:
\begin{equation}
    \mathbf{c} = \mathcal{F}(\mathbf{x}^{\mathrm{tgt}}_{\mathrm{cond}}), \quad
    \mathbf{g} = \mathcal{S}(\mathbf{x}^{\mathrm{tgt}}_{\mathrm{cond}}),
\end{equation}
where $\mathcal{F}(\cdot)$ denotes the acoustic feature extractor and $\mathcal{S}(\cdot)$ denotes the speaker encoder.

The acoustic converter performs voice conversion directly in the codec latent space:
\begin{equation}
    \hat{\mathbf{z}} = f(\mathbf{z}, \mathbf{c}, \mathbf{g}),
\end{equation}
where $\hat{\mathbf{z}}$ is the converted codec latent sequence.
Finally, the codec decoder reconstructs the waveform:
\begin{equation}
    \hat{\mathbf{x}} = \mathcal{D}(\hat{\mathbf{z}}).
\end{equation}

As illustrated in Figure~\ref{fig:model_overview}, we first encode the input waveform into codec latent representations using a pretrained speech encoder.
Voice conversion is then performed on these latent representations by the acoustic converter, before the converted latents are decoded back into waveform.
This formulation allows us to cast voice conversion as a transformation in a pretrained codec latent space, rather than directly generating waveform samples.

\subsection{Codec Instantiation with SAC}
\label{subsec:sac}

In this work, the abstract speech encoder introduced in Section~\ref{subsec:overview} is instantiated by the frontend of the pretrained SAC codec~\cite{Chen2025SAC}.
Concretely, it includes the semantic encoder, VQ and adapter, the acoustic encoder and VQ, and the prenet.

For the zero-shot streaming voice conversion task considered in this work, SAC offers two particularly useful properties.

First, SAC provides a strong latent frontend for high-quality speech reconstruction.
Its dual-stream design combines semantic and acoustic information, yielding latent representations that preserve both linguistic content and fine-grained acoustic characteristics.
This makes SAC a suitable backbone for voice conversion, where the model must preserve source content while adapting speaker-related acoustic traits.

Second, SAC is trained on short speech segments, which makes it naturally compatible with chunkwise processing at inference time.
Although the codec itself is not strictly causal, this segment-based training paradigm is favorable for low-latency chunkwise conversion with overlap smoothing, as described in Section~\ref{subsec:streaming}.

At the same time, from the perspective of voice conversion, we do not regard SAC as providing perfectly disentangled semantic and acoustic representations.
Although SAC organizes speech information into semantic and acoustic pathways, its acoustic branch is still optimized for accurate waveform reconstruction and therefore may retain content-related information in addition to speaker and acoustic traits.
In this sense, SAC is better viewed as a high-quality codec frontend than as a fully disentangled tokenizer.
This is also different from works that explicitly strengthen semantic--acoustic separation through dedicated supervision and recombination-style training, such as DSA-Tokenizer~\cite{zhang2026dsatokenizerdisentangledsemanticacoustictokenization}.

For this reason, we do not perform conversion directly on the acoustic codes of a single pathway.
Instead, we perform conversion after the prenet, where information from both pathways has already been integrated into a unified latent space.
This stage corresponds to the output of the speech encoder in Figure~\ref{fig:model_overview}.
Operating on this unified representation allows the converter to jointly model the content that should be preserved and the acoustic characteristics that should be transformed, which is better aligned with the goal of voice conversion.

We build upon the pretrained codec without redesigning its architecture.
During training, the codec encoder-side components are kept frozen so that the latent space remains stable and learning is focused on the conversion module.
In practice, this includes the semantic and acoustic encoding pathways, together with their quantization modules.
The proposed acoustic converter (detailed next in Section~\ref{subsec:converter}) is then optimized to transform the source codec latents conditioned on the target-side reference speech, while the decoder reconstructs the converted waveform from the transformed latents.

\subsection{Acoustic Converter}
\label{subsec:converter}

\begin{figure}
    \centering
    \includegraphics[width=0.8\linewidth]{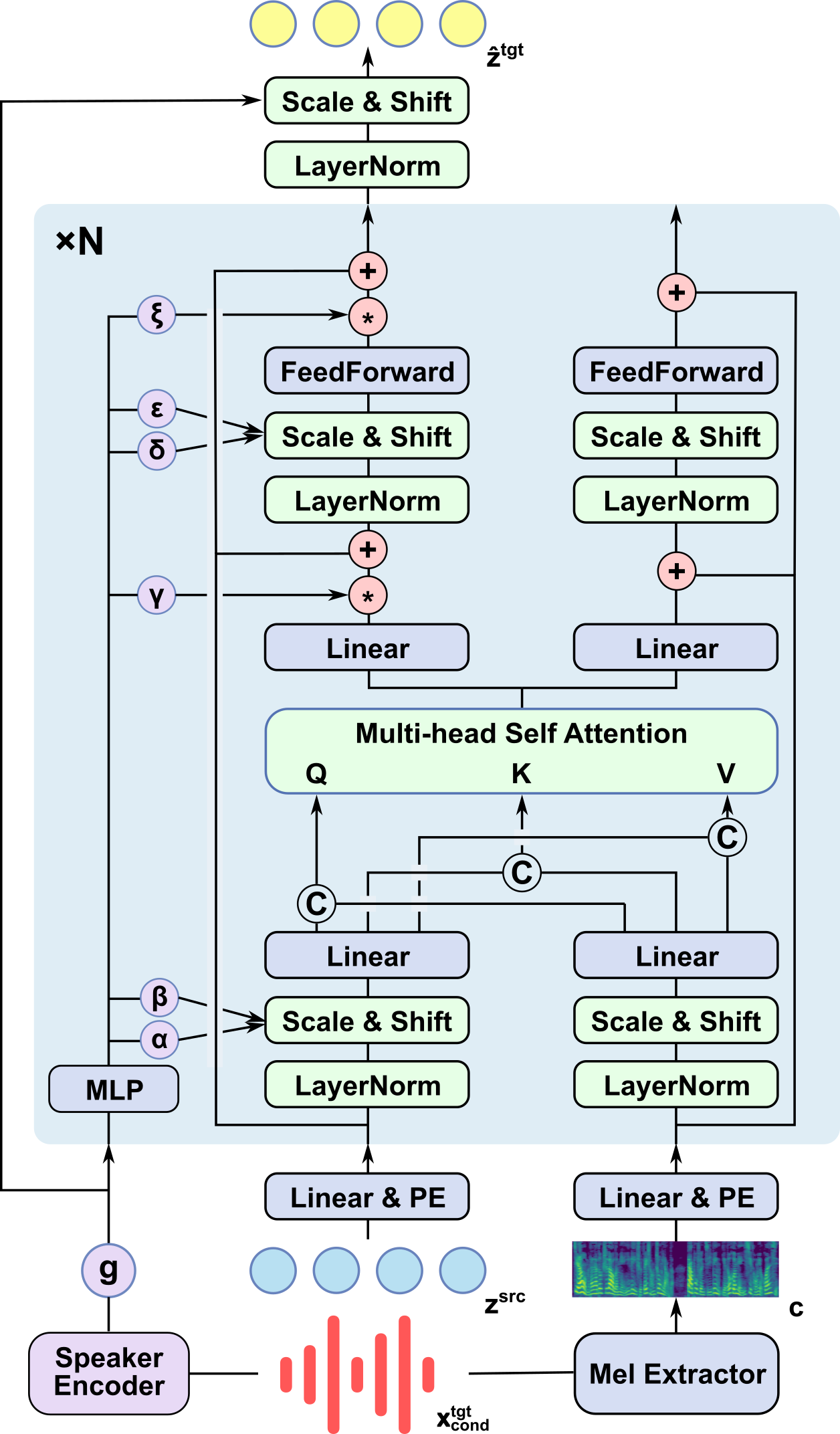}
    \caption{Architecture of the acoustic converter. Source codec latents and the mel sequence extracted from $\mathbf{x}^{\mathrm{tgt}}_{\mathrm{cond}}$ are jointly updated through shared attention, while the utterance-level speaker embedding modulates each block through adaptive normalization.}
    \label{fig:converter}
    \Description{The figure illustrates a stacked transformer-like architecture with N repeated blocks. Each block contains two parallel processing streams corresponding to different modalities. Both streams include LayerNorm followed by scale-and-shift modulation, linear projections, multi-head self-attention, and feed-forward networks with residual connections. The attention module receives queries, keys, and values from both streams, with concatenation points indicated by “C”. Global conditioning parameters (gamma, beta, and others) are generated from a speaker embedding through an MLP and injected via scale-and-shift operations. The lower part shows separate input embeddings with positional encoding for acoustic tokens and mel-spectrogram features, as well as a speaker encoder that produces the global conditioning vector.}
\end{figure}

The acoustic converter is the core component of the proposed system.
It transforms source codec latents toward the target speaker characteristics while preserving the linguistic content.
An overview of its architecture is shown in Figure~\ref{fig:converter}.

\subsubsection{Design Overview}

Our goal is to incorporate two complementary forms of conditioning into the conversion process: 
a frame-level acoustic condition and an utterance-level speaker representation.
The former provides detailed, time-varying acoustic patterns of the target speaker, while the latter provides a stable global description of speaker identity.

A key challenge is that these signals are defined in different representation spaces.
The source is represented in codec latent space, while the frame-level condition is derived from mel-spectrograms.
These representations cannot be directly combined at the input level.
To address this, we design the converter as a two-branch architecture with separate input projections for the source and the condition.
Inspired by recent multimodal transformer~\cite{StabilityAI2024SD3} architectures, this design allows heterogeneous inputs to be processed independently before enabling interaction between them.

During training, the mel sequence fed to the conditioning branch is extracted from $\mathbf{x}^{\mathrm{tgt}}_{\mathrm{cond}}$, i.e., the target utterance after removing the selected target segment.
This prevents trivial copying and keeps the conditioning setup consistent with the training pipeline described in Section~\ref{subsec:training}.

\subsubsection{Interaction Between Source and Condition}

After projection, the source and condition sequences are processed jointly through a stack of transformer blocks.
In each block, the two sequences are concatenated for the attention operation, enabling information exchange between them.

Importantly, the condition is not treated as a fixed context.
Both the source and condition representations are updated across layers.
This allows the model to progressively refine both representations, rather than only conditioning the source on a static reference.

In the context of voice conversion, this behavior is particularly important.
The source latent carries the linguistic content to be preserved, while the acoustic condition provides detailed spectral patterns of the target speaker.
By jointly updating both representations, the model can progressively align the source content with the target acoustic characteristics.

\subsubsection{Global Speaker Conditioning}

In addition to the frame-level acoustic condition, we incorporate an utterance-level speaker embedding as a global conditioning signal.
This embedding is injected into the network through adaptive normalization layers, which modulate the hidden representations across layers.

The global speaker representation complements the frame-level condition.
While the mel-spectrogram provides local acoustic details, the speaker embedding provides a consistent identity signal over the entire utterance.
This helps stabilize the conversion and improves speaker consistency.

\subsection{Training Strategy}
\label{subsec:training}

\begin{figure}
    \centering
    \includegraphics[width=0.8\linewidth]{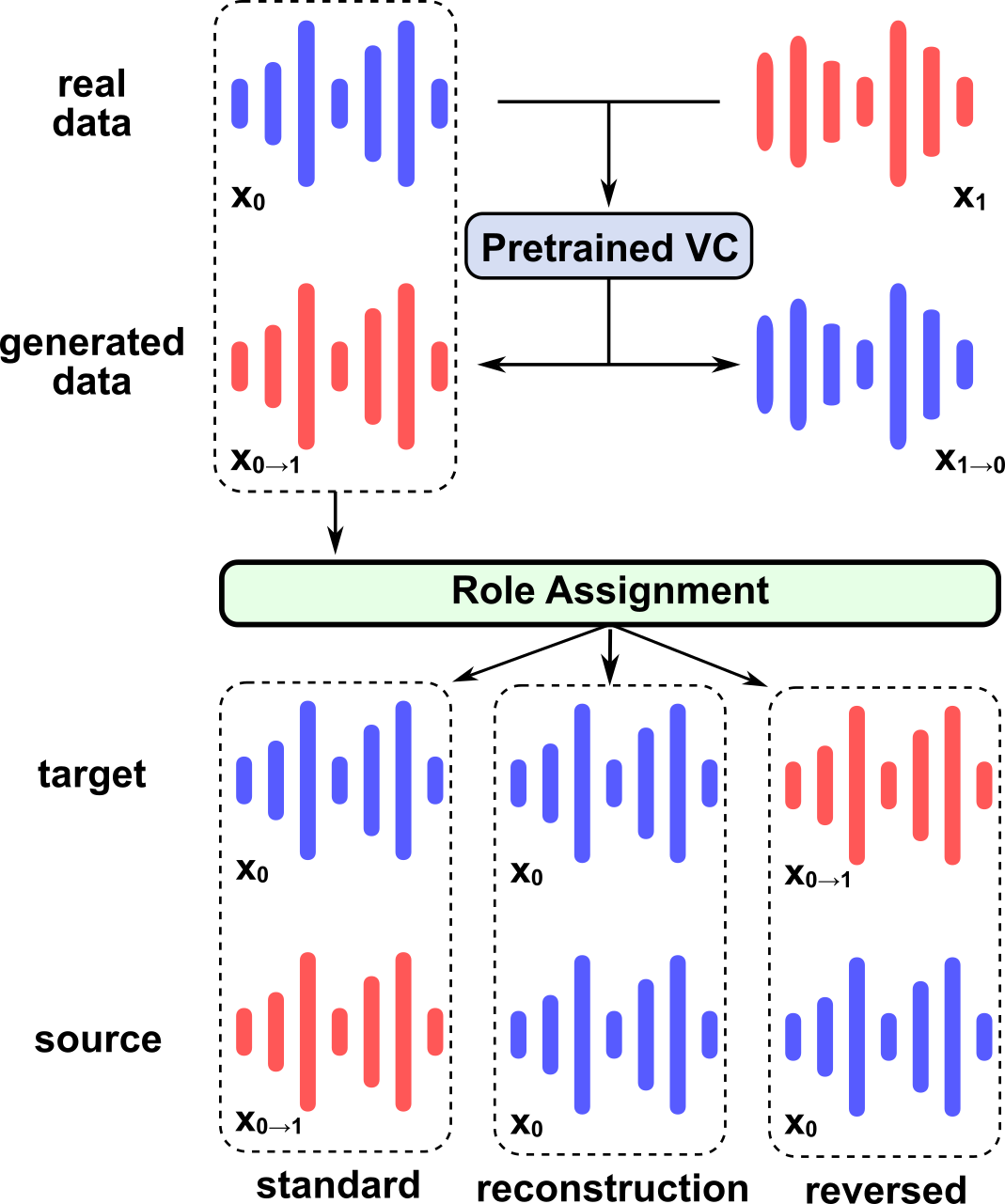}
    \caption{Training data construction and role assignment. A pretrained VC model generates paired utterances from real speech, which are then used to form source-target pairs under standard, reconstruction, and reversed modes.}
    \label{fig:data_pipeline}
    \Description{The diagram shows the training data construction process. Real utterances x0 and x1 are first used to generate converted samples x0 to 1 and x1 to 0 with a pretrained voice conversion model. One real utterance and one generated utterance with matched content are then passed to a role assignment stage. Three role configurations are illustrated: standard, where the generated utterance is used as the source and the real utterance is used as the target; reconstruction, where both source and target are taken from the same real utterance; and reversed, where the real utterance is used as the source and the generated utterance is used as the target.}
\end{figure}

Our training strategy is based on paired data with matched linguistic content and different speaker characteristics, together with flexible role assignment between source and target.
An overview is shown in Figure~\ref{fig:data_pipeline}.

\subsubsection{Data Construction}
\label{subsubsec:data_construction}

For each randomly paired utterance pair $(\mathbf{x}_0, \mathbf{x}_1)$, generated samples are constructed in both directions, i.e., $\mathbf{x}_{0\rightarrow1}$ and $\mathbf{x}_{1\rightarrow0}$.
Each generated sample preserves the linguistic content of one utterance while adopting the speaker characteristics of the other.
This process is performed using a pretrained offline voice conversion model.

As a result, each training pair consists of two utterances with matched linguistic content but different speaker characteristics, one real and one generated.
Such pairs provide suitable supervision for voice conversion, where content and speaker identity are disentangled at the data level.

During training, both source and target waveforms are cropped into short segments.
For the frame-level conditioning branch, we remove the selected target segment from the original target utterance and use the remaining part, denoted as $\mathbf{x}^{\mathrm{tgt}}_{\mathrm{cond}}$, to extract the mel-spectrogram condition.
This prevents the conditioning branch from directly accessing the target segment and avoids trivial copying.

\subsubsection{Role Assignment}
\label{subsubsec:role_assignment}

Given a paired example consisting of two utterances with the same content but different speaker characteristics, we consider different assignments of source and target roles.

A standard choice is to use the generated utterance as the source and the corresponding real utterance as the target.
For example, given the pair $(\mathbf{x}_{0\rightarrow1}, \mathbf{x}_0)$, the standard assignment uses $\mathbf{x}_{0\rightarrow1}$ as the source and $\mathbf{x}_0$ as the target.

In addition, we also use the reversed assignment, where the real utterance is used as the source and the generated utterance is used as the target.
For the same example, the reversed assignment uses $\mathbf{x}_0$ as the source and $\mathbf{x}_{0\rightarrow1}$ as the target.

We further include a self-reconstruction assignment, where the source and target are taken from the same real utterance, e.g., $(\mathbf{x}_0, \mathbf{x}_0)$.

During training, each paired example is assigned to one of the three modes---standard, self-reconstruction, or reversed---by random sampling according to probabilities $p_{\text{std}}$, $p_{\text{recon}}$, and $p_{\text{rev}}$.

These assignments do not change the semantic correspondence of the pair, but expose the model to a more diverse input-output distribution.
In particular, both the encoder and decoder observe real and generated speech during training, which improves robustness and reduces distribution mismatch.

\subsubsection{Training Objectives}
\label{subsubsec:training_objective}

The training objective follows the original SAC formulation, with the main difference being the definition of the prediction target.

Given a selected target segment under the role assignment above, the model is trained to reconstruct this target from the corresponding source and conditioning signals.
Therefore, all losses are computed with respect to the chosen target segment.

We use the same loss terms as in SAC, including the semantic MSE loss, mel reconstruction loss, speaker similarity MSE loss, and adversarial discriminator loss.
Since the vector quantization modules are frozen in our setting, the VQ-related loss is removed.

\subsection{Chunkwise Streaming Inference}
\label{subsec:streaming}

\begin{figure}
    \centering
    \includegraphics[width=\linewidth]{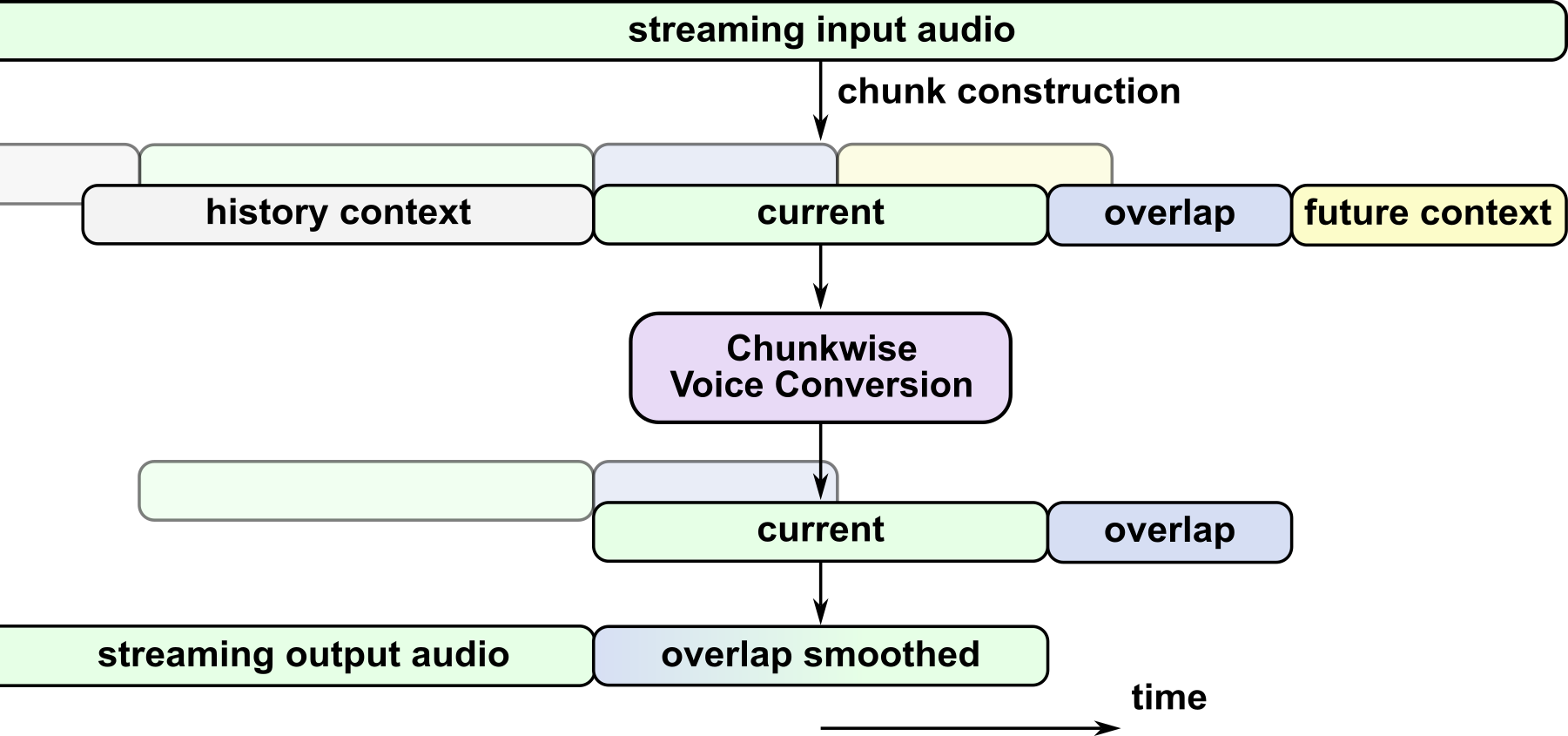}
    \caption{Chunkwise streaming inference of X-VC. Each step uses a fixed processing window composed of history, current, overlap, and future regions. Only the current region is emitted as valid output, while the front part of the current chunk is blended with the overlap retained from the previous step to ensure continuity.}
    \label{fig:streaming}
    \Description{The diagram shows a streaming processing timeline for input and output audio. The input audio is segmented into multiple regions labeled history context, current segment, overlap, and future context. These segments are fed into a chunkwise voice conversion module. The module processes the current segment while utilizing past and limited future information. The output is generated in chunks corresponding to the current segments, with adjacent outputs overlapping in time. The overlapping regions are merged using smoothing to produce a continuous output waveform.}
\end{figure}

Although the underlying codec is not strictly causal, its segment-based training paradigm makes it amenable to chunkwise inference.
To enable low-latency voice conversion, chunkwise inference is applied only in streaming scenarios, as illustrated in Figure~\ref{fig:streaming}.

During streaming inference, the conditioning signals are precomputed from the reference speech.
Specifically, the frame-level acoustic features and the utterance-level speaker embedding are extracted once before streaming begins, and are reused for all subsequent chunks.
Therefore, only the source speech is processed in a streaming manner.
The input speech is processed in consecutive chunks.
At each step, a fixed-length window is constructed, consisting of four parts: a history region, a current region, an overlap region for boundary smoothing, and an optional future context.
The total window length is aligned with the training segment, ensuring consistency between training and inference.
The entire window is fed into the codec and acoustic converter, while only the current region is retained as valid output.
The surrounding context is introduced to improve local continuity but is not directly emitted.
This design allows the model to leverage sufficient temporal context while maintaining low latency.

To ensure smooth transitions across chunk boundaries, overlap smoothing is applied.
Specifically, the overlap retained from the previous step is blended with the front part of the current chunk using a cosine cross-fade, reducing discontinuities in the reconstructed waveform.

The algorithmic latency is determined by both the required input context and the computation time.
We decompose it as:
\begin{equation}
T_{\text{latency}} = T_{\text{model}} + T_{\text{compute}},
\label{eq:latency}
\end{equation}
where $T_{\text{model}}$ denotes the amount of input context required before producing valid output, and $T_{\text{compute}}$ denotes the runtime of the model.
In this system,
\begin{equation}
T_{\text{model}} = T_{\text{current}} + T_{\text{overlap}} + T_{\text{future}},
\label{eq:model_latency}
\end{equation}
\begin{equation}
T_{\text{compute}} = T_{\text{enc}} + T_{\text{convert}} + T_{\text{dec}}.
\label{eq:compute}
\end{equation}
Here, $T_{\text{compute}}$ includes all online processing steps in the pipeline, including encoding, acoustic conversion, and decoding, corresponding to an end-to-end definition of streaming inference.
Conditioning features derived from the reference speech are computed beforehand and thus excluded.
Overall, this chunkwise strategy approximates streaming behavior and maintains low latency without requiring a fully causal model, all while preserving the reconstruction quality of the pretrained codec.

\begin{table*}
\caption{Streaming zero-shot voice conversion results on Seed-TTS-Eval. Best values are shown in bold. $\uparrow$ and $\downarrow$ indicate higher or lower values are better. Because the test set is non-parallel, only GT WER is reported, using the source utterances as content references; GT SIM and GT UTMOS are omitted as they are not meaningful in this setting.}
\label{tab:main_result_streaming}
\centering
{
\begin{tabular}{lrrccccccrr}
\toprule
\textbf{Model} & \multicolumn{2}{c}{\textbf{\#Param.}} & 
\multicolumn{3}{c}{\textbf{Seed-TTS \textit{test-en}}} & 
\multicolumn{3}{c}{\textbf{Seed-TTS \textit{test-zh}}} & \multicolumn{2}{c}{\textbf{Latency (ms) $\downarrow$}} \\
\cmidrule(lr){2-3}\cmidrule(lr){4-6} \cmidrule(lr){7-9} \cmidrule(lr){10-11}
& \textbf{Total} & \textbf{Converter} & \textbf{WER (\%) $\downarrow$} & \textbf{SIM $\uparrow$} & \textbf{UTMOS $\uparrow$} & \textbf{WER (\%) $\downarrow$} & \textbf{SIM $\uparrow$} & \textbf{UTMOS $\uparrow$} & {$T_{\text{model}}$} & $T_{\text{compute}}$ \\
\midrule
Ground Truth                  & -    & -     & 1.96    & -     & -    & 1.33 & -     & -    & - & - \\
\midrule
Seed-VC$_\textit{tiny}$       & 262M & 27M   & 3.31    & 0.40  & 2.97 & 3.36 & 0.60  & \textbf{2.46} & 380 & 120.75 \\
MeanVC                        & 47M  & 14M   & -       & -     & -    & 4.89 & \textbf{0.72}  & 2.22 & 250 & \textbf{32.12} \\
\textbf{X-VC (Ours)}                  & 539M & 44M   & \textbf{3.14}    & \textbf{0.62}  & \textbf{3.07} & \textbf{2.65} & \textbf{0.72}  & 2.35 & 240 & 58.17 \\
\bottomrule
\end{tabular}
}
\end{table*}

\begin{table}[t]
\centering
\caption{Subjective speaker similarity (SMOS $\uparrow$) with 95\% \\ confidence intervals computed from the t-test.}
\label{tab:subjective_sim}
\begin{tabular}{lcc}
\toprule
\textbf{Model} & \textbf{Seed-TTS \textit{test-en}} & \textbf{Seed-TTS \textit{test-zh}} \\
\midrule
Seed-VC   & 3.56 $\pm$ 0.12 & 3.38 $\pm$ 0.17 \\
MeanVC    & - & 3.44 $\pm$ 0.15                \\
\textbf{X-VC (Ours)} & \textbf{3.98} $\pm$ \textbf{0.10} & \textbf{3.89} $\pm$ \textbf{0.13} \\
\bottomrule
\end{tabular}
\end{table}
\section{Experimental Setup}

\subsection{Model Configuration}

\subsubsection{Codec, Feature Extraction, and Speaker Encoder}

A pretrained SAC codec~\cite{Chen2025SAC} is adopted, using its 16\,kHz 62.5\,Hz configuration.
The speech encoder encodes speech into codec latent representations with a dimensionality of 1024.
For frame-level conditioning, mel-spectrograms are extracted from the reference speech using a standard mel extractor with 128 mel bins, which defines the sequence-level conditioning dimension.
For utterance-level conditioning, a pretrained speaker encoder based on ERes2Net~\cite{Chen2023ERes2Net} is used to produce 192-dimensional speaker embeddings.

\subsubsection{Converter}

The acoustic converter follows the architecture described in Section~\ref{subsec:converter}.
It is implemented as a stack of transformer blocks that jointly model source codec latents and frame-level acoustic conditions.
The input dimensionalities are determined by the upstream components:
the codec latent has 1024 channels, the frame-level condition has 128 channels (mel-spectrogram), and the utterance-level speaker embedding has dimension 192.
A hidden dimension of 512 is used, with 6 transformer layers and 8 attention heads, where each head has a dimension of 64.
A feed-forward expansion ratio of 4 is adopted in each block. 

\subsubsection{Training Details}

Training is performed using the strategy described in Section~\ref{subsec:training}.
$(p_{\text{std}},\; p_{\text{recon}},\; p_{\text{rev}})$ are set to $(0.4, 0.2, 0.4)$.
The model is trained for 446k steps on 8 NVIDIA H200 GPUs, with a per-GPU batch size of 24.
The AdamW optimizer is used with a learning rate of $1\times10^{-4}$ and $\beta=(0.8, 0.9)$.
An exponential learning rate scheduler is applied with a decay factor of $0.999996$ and a minimum learning rate of $10^{-6}$. 
Gradient clipping is applied with a maximum norm of 5, and exponential moving average (EMA) is used to stabilize training. 

\subsection{Datasets}

\subsubsection{Training Dataset}

The model is trained on a mixture of the Emilia dataset~\cite{He2024Emilia} and the LibriTTS dataset~\cite{Zen2019LibriTTS}.
For the Emilia dataset, data filtering is applied based on DNSMOS~\cite{Reddy2020DNSMOS}, and only samples with DNSMOS scores higher than 3.45 are retained to ensure data quality.
In total, approximately 10,000 hours of English and Chinese speech data are collected.
Based on these data, additional paired samples are generated using a pretrained Seed-VC$_{\textit{small}}$ model, following the procedure described in Section~\ref{subsubsec:data_construction}.
This results in approximately 20,000 hours of generated data. 
Following SAC, each audio sample is resampled to 16 kHz and randomly cropped into 2.4-second segments during training. 

\subsubsection{Evaluation Dataset}

We evaluate our model on the Seed-TTS-Eval benchmark~\cite{Anastassiou2024SeedTTS}, which offers non-parallel pairs for zero-shot voice conversion.
The evaluation includes English (\textit{test-en}) and Chinese (\textit{test-zh}) settings.
In addition, we construct a cross-lingual evaluation setting based on the Seed-TTS-Eval dataset.
Specifically, we select the first 1,000 utterances from the English and Chinese subsets, and form cross-lingual pairs by exchanging the source and reference utterances across languages.
This results in two subsets: English-to-Chinese and Chinese-to-English, which evaluate the ability of the model to transfer speaker characteristics across languages while preserving the original linguistic content.

\subsection{Baseline Models}

We compare the proposed method with representative zero-shot voice conversion models trained on the Emilia dataset.

Seed-VC~\cite{Liu2024SeedVC} is a diffusion-based voice conversion framework.
Following the guidelines in its official repository\footnote{\url{https://github.com/Plachtaa/seed-vc}}, we evaluate Seed-VC$_{\text{small}}$ in the offline setting and Seed-VC$_{\text{tiny}}$ in both streaming and offline settings.
MeanVC~\cite{Ma2025MeanVC} is a flow-based voice conversion model built upon the MeanFlow framework.
We use its released 200\,ms chunk version\footnote{\url{https://github.com/ASLP-lab/MeanVC}} for both streaming and offline settings.
We note that its released offline inference script is also chunkwise rather than full-utterance, which should be taken into account when interpreting its offline RTF. Also, MeanVC is omitted from the English-source evaluations as its content encoder module only supports Chinese.

\subsection{Evaluation Metrics}

We evaluate performance across both generation quality and inference efficiency. To assess speech quality, we use Word Error Rate (WER) to measure linguistic preservation, employing Whisper-large-v3 \cite{Radford2023Whisper} for English and Paraformer-zh \cite{Gao2022Paraformer} for Chinese. Speaker identity consistency is evaluated via Speaker Similarity (SIM), computed as the cosine similarity between speaker embeddings extracted by a WavLM-based model \cite{Chen2022WavLM}, while UTMOS \cite{Saeki2022UTMOS} is utilized to objectively estimate speech naturalness without human annotations. For inference efficiency, all metrics are measured end-to-end on a single NVIDIA RTX 3090 GPU. Streaming latency is decomposed into model-induced latency ($T_{\text{model}}$), determined by the required current, overlap, and future contexts, and computation time ($T_{\text{compute}}$), which represents the average runtime per chunk including encoding, acoustic conversion, and decoding. Finally, for offline inference, we report the real-time factor (RTF) as the ratio of the total processing time to the input audio duration.

In addition to objective metrics, we perform a subjective evaluation to assess speaker similarity. We conduct a speaker similarity mean opinion score (SMOS) test with 15 listeners, who are asked to rate the speaker similarity between the converted speech and the target reference on a 5-point scale (1: completely different, 5: identical speaker). All samples are presented in random order to avoid bias.


\begin{table*}
\caption{Offline zero-shot voice conversion results on Seed-TTS-Eval.}
\label{tab:main_result_offline}
\centering
{
\begin{tabular}{lrrccccccc}
\toprule
\textbf{Model} & \multicolumn{2}{c}{\textbf{\#Param.}} & 
\multicolumn{3}{c}{\textbf{Seed-TTS \textit{test-en}}} & 
\multicolumn{3}{c}{\textbf{Seed-TTS \textit{test-zh}}} & \textbf{RTF $\downarrow$} \\
\cmidrule(lr){2-3}\cmidrule(lr){4-6} \cmidrule(lr){7-9}
& \textbf{Total} & \textbf{Converter} & \textbf{WER (\%) $\downarrow$} & \textbf{SIM $\uparrow$} & \textbf{UTMOS $\uparrow$} & \textbf{WER (\%) $\downarrow$} & \textbf{SIM $\uparrow$} & \textbf{UTMOS $\uparrow$} &  \\
\midrule
Ground Truth                  & -    & -     & 1.96    & -     & -    & 1.33 & -     & - & - \\
\midrule
Seed-VC$_\textit{small}$      & 296M & 98M   & 2.57    & 0.56  & 3.37 & 2.52 & \textbf{0.73}  & 2.73 & 0.161 \\
Seed-VC$_\textit{tiny}$       & 262M & 27M   & \textbf{2.24}    & 0.41  & \textbf{3.67} & \textbf{1.79} & 0.60  & \textbf{3.08} & 0.069 \\
MeanVC                        & 47M  & 14M   & -       & -     & -    & 3.89 & \textbf{0.73}  & 2.81 & 0.094 \\
\textbf{X-VC (Ours)}                   & 539M & 44M   & 2.83    & \textbf{0.63}  & 3.31 & 1.99 & \textbf{0.73}  & 2.69 & \textbf{0.014} \\
\bottomrule
\end{tabular}
}
\end{table*}

\begin{table}
\caption{Offline cross-lingual results on Seed-TTS-Eval.}
\label{tab:cross_lingual}
\centering
{
\begin{tabular}{lcccc}
\toprule
\textbf{Model} &  
\multicolumn{2}{c}{\textbf{English-to-Chinese}} & 
\multicolumn{2}{c}{\textbf{Chinese-to-English}} \\
\cmidrule(lr){2-3} \cmidrule(lr){4-5}
& \textbf{WER (\%)$\downarrow$} & \textbf{SIM $\uparrow$}  & \textbf{WER (\%)$\downarrow$} & \textbf{SIM $\uparrow$}  \\
\midrule
Ground Truth                         & 1.91    & -     & 1.14 & -      \\
\midrule
Seed-VC$_\textit{small}$             & 2.82    & \textbf{0.52}  & 2.36 & 0.42   \\
Seed-VC$_\textit{tiny}$              & \textbf{2.23}    & 0.34  & \textbf{1.70} & 0.29  \\
MeanVC                               & -       & -     & 3.89 & 0.42  \\
\textbf{X-VC (Ours)}                          & 2.67    & \textbf{0.52}  & 2.15 & \textbf{0.49}  \\
\bottomrule
\end{tabular}
}
\end{table}

\begin{table}
\caption{Offline ablation results on Seed-TTS \textit{test-zh}. \\
All models are trained for 300k steps for fair comparison.}
\label{tab:ablation}
\centering
{
\begin{tabular}{lcc}
\toprule
\textbf{Model} & \textbf{WER (\%)$\downarrow$} & \textbf{SIM $\uparrow$} \\
\midrule
X-VC                                 & \textbf{2.02} & \textbf{0.72} \\
\midrule
\textit{Condition Modeling} \\ 
{ } { } w/o updating frame-level condition $\mathbf{c}$    & 2.15 & 0.66  \\ 
{ } { } w/o utterance-level condition $\mathbf{g}$         & 2.20 & 0.61  \\ 
\midrule
\textit{Data Construction} \\ 
{ } { } standard only ($p_{\text{std}} = 1.0$)                                 & 2.31 & \textbf{0.72}  \\ 
{ } { } reversed only ($p_{\text{rev}} = 1.0$)                                 & 2.14 & 0.71  \\ 
{ } { } w/o reconstruction ($p_{\text{std}} = p_{\text{rev}} = 0.5$)           & 2.14 & \textbf{0.72}  \\ 

\bottomrule
\end{tabular}
}
\end{table}

\section{Results and Discussion}

\subsection{Streaming Performance}

Table~\ref{tab:main_result_streaming} reports the streaming zero-shot voice conversion results on Seed-TTS-Eval. 
Overall, X-VC provides the strongest streaming performance among the compared systems.
It achieves the best WER on both \textit{test-en} and \textit{test-zh}, indicating more reliable content preservation under online constraints.
At the same time, its SIM is the best on English and tied for the best on Chinese, while its UTMOS remains competitive and is the best on English.
Taken together, these results suggest that X-VC offers a better overall balance of intelligibility, speaker similarity, and perceptual quality than the streaming baselines.

In addition to objective metrics, the subjective speaker similarity results in Table~\ref{tab:subjective_sim} are also examined. 
Results show that X-VC achieves higher SMOS scores compared to the baseline systems, indicating better perceptual speaker similarity. 
Meanwhile, X-VC exhibits lower variance in listener ratings, suggesting more consistent conversion quality across samples. 
These observations are consistent with the objective SIM results.

In terms of latency, X-VC has a model-induced latency $T_{\text{model}}$ of 240\,ms, with 120\,ms for $T_{\text{current}}$, 20\,ms for $T_{\text{overlap}}$, and 100\,ms for $T_{\text{future}}$.
Its computation latency $T_{\text{compute}}$ is higher than that of MeanVC, but remains lower than that of Seed-VC$_\textit{tiny}$.
This comparison should be interpreted together with the streaming context design.
To be consistent with the 2.4-second segment setting used in training, each streaming step of X-VC uses a 2.4-second processing window, which is substantially larger than those of the compared baselines, especially in terms of history context.
This larger window also provides stronger contextual support for chunkwise streaming inference, leading to more stable conversion across chunk boundaries.
Even with this larger context, X-VC remains streamable, indicating that the proposed one-step codec-space conversion is computationally practical for low-latency inference.

\subsection{Offline Performance}

Table~\ref{tab:main_result_offline} shows the offline zero-shot voice conversion results.
Under the offline setting, X-VC remains competitive in quality while exhibiting a clear advantage in inference efficiency.
In particular, it achieves the best SIM on English and tied-best SIM on Chinese, while its WER remains competitive, especially on Chinese.
Its UTMOS is not always the best, but stays in the same overall range as the strongest baselines.
These results indicate that X-VC is particularly effective at speaker similarity preservation while maintaining solid intelligibility and naturalness.

A notable advantage of X-VC is efficiency.
It achieves an RTF of 0.014, which is substantially lower than all compared baselines.
This is consistent with the one-step nature of the proposed method: once the codec latents are extracted, conversion is performed directly in latent space without iterative generation.
We also note that the released offline script of MeanVC performs chunkwise inference rather than full-utterance inference, which partly explains its relatively low RTF.
Nevertheless, X-VC remains considerably faster.
Overall, the offline results show that X-VC provides a strong quality--efficiency trade-off.

\subsection{Cross-lingual Evaluation}

Table~\ref{tab:cross_lingual} presents the cross-lingual results.
X-VC remains competitive in both language-mismatched settings.
For English-to-Chinese conversion, it matches the best SIM and achieves WER close to the best baseline.
For Chinese-to-English conversion, it achieves the best SIM while maintaining competitive WER.
These results are consistent with the same-language experiments: X-VC is particularly strong at preserving target speaker characteristics, while remaining robust in content preservation under cross-lingual mismatch.

\subsection{Ablation Studies}

Table~\ref{tab:ablation} reports two groups of ablations on Seed-TTS \textit{test-zh}.

For condition modeling, removing either the utterance-level condition $\mathbf{g}$ or the updating of frame-level condition $\mathbf{c}$ degrades both WER and SIM, indicating that the two conditioning signals are complementary and both important for content preservation and speaker similarity.

For data construction, the ablations mainly affect WER while leaving SIM largely unchanged. This suggests that the multi-mode role-assignment strategy primarily improves robustness and intelligibility, with more limited impact on speaker similarity.

Overall, the ablations confirm the importance of both the dual-conditioning converter and the proposed data-construction strategy.
\section{Conclusion}

This paper presented X-VC, a zero-shot streaming voice conversion system that performs one-step conversion in the latent space of the pretrained SAC codec.
The proposed system combines a dual-conditioning acoustic converter, generated-pair training with flexible role assignment, and chunkwise streaming inference with overlap smoothing.
Experiments on Seed-TTS-Eval demonstrate that X-VC achieves the best streaming WER in both English and Chinese, strong speaker similarity in both same-language and cross-lingual settings, and substantially lower offline RTF than the compared baselines.
Ablation studies further confirm the importance of jointly updating the frame-level condition stream, incorporating utterance-level global speaker conditioning, and using role assignment during training.

Overall, the results suggest that codec-space one-step conversion is a promising direction for practical zero-shot voice conversion, especially when both quality and latency are important.

\clearpage

\bibliographystyle{ACM-Reference-Format}
\bibliography{myref}

\end{document}